\documentclass{fundam}

\publyear{22}
\papernumber{2102}
\volume{185}
\issue{1}

\usepackage{url} % takes care of hyperlinks, preferred over hyperref
\usepackage[ruled,lined]{algorithm2e}% provides Algorithm environment
\usepackage{graphicx}% allows for inclusion of graphic files (figures)
\usepackage{tikz}
\usetikzlibrary{automata, positioning, arrows}

\usepackage{mymacros}

\begin{document}

\title{Homomorphisms on a product of free convex semilattices}

\address{ana.sokolova@cs.uni-salzburg.at}

\author{Ana Sokolova\\
Department of Computer Science\\
University of Salzburg\\ 
Salzburg, Austria\\
ana.sokolova@cs.uni-salzburg.at
\and Harald Woracek\\
Institute for Analysis and Scientific Computing\\
TU Wien \\
Vienna, Austria\\
harald.woracek@tuwien.ac.at
} 

\maketitle

\runninghead{A.Sokolova, H.Woracek}{Homomorphisms on a product of free convex semilattices}

\begin{abstract}
	Convex semilattices form an equational class that combines convex algebras and semilattices by a distributive law.
	For a set $I$ let $\mc FI$ be the free convex semilattice with basis $I$. A product $\mc FI\times\mc FJ$ is generated by
	$I\times J$, but is not free. Hence, not every map from $I\times J$ into an arbitrary convex semilattice $X$ 
	extends uniquely to a homomorphism from $\mc FI\times\mc FJ$ to $X$. 
	We investigate a condition on a map $\varphi$ of $I\times J$ into some convex semilattice $X$
	that is always necessary and often sufficient for $\varphi$ to have such an extension to a homomorphism $\Phi$ of 
	$\mc FI\times\mc FJ$ into $X$.
\end{abstract}

\begin{keywords}
	convex semilattice, homomorphisms, product of free
\end{keywords}

%%%%%%%%%%%%%%%%%%%%%%%%%%%%%%%%%%%%%%%%%%%%%%%%%%%%%%%%%%%%%%%%%%%%%%

%
%
%
\section{Introduction}

A convex semilattice is a combination of a convex algebra and a semilattice, imposing a distributive law that rules the 
interplay between these structures (we refer to \Cref{H19} for all definitions). 
This kind of algebras has a more abstract background, namely as the Eilenberg-Moore algebras of a certain monad, and has
recently attracted attention in theoretical computer science, in the context of traces of transition systems with nondeterminism and
probability, see \cite{bonchi.sokolova.vignudelli:2022}, see also \cite{mio.vignudelli:2020,mio.sarkis.vignudelli:2021} for
quantitative extensions.
From a different viewpoint, algebras whose operations distribute over a semilattice operation have been studied 
e.g.\ in \cite{romanowska.smith:1985,smith:2011,hyndman.nation.nishida:2016}.

In the present paper we deal with a very concrete question about convex semilattices. 
The question itself can be formulated in any equational class.
\begin{itemize}
\item[{\bf(Q)}] {\it
	Let $\bb V$ be an equational class. For a nonempty set $I$ denote by $\mc FI$ the free algebra in $\bb V$ with basis $I$. 
	Given two nonempty sets $I,J$, an algebra $X$ in $\bb V$, and a map $\varphi\DF I\times J\to X$, does there exist a 
	homomorphism $\Phi\DF\mc FI\times\mc FJ\to X$ with $\Phi|_{I\times J}=\varphi$ ?
	}
	\begin{equation}
	\label{H1}
		\begin{tikzcd}[column sep=large]
			I\times J \arrow[r,"\subseteq",tail] \arrow[rd,swap,"\varphi"]
			& \mc FI\times\mc FJ \arrow[d,dashed,"\Phi"] 
			\\
			& X
		\end{tikzcd}
	\end{equation}
\end{itemize}
Recall that the free algebra $\mc FI$ is characterised by the property that any map $I\to X$ into some algebra $X\in\bb V$ 
extends uniquely to a homomorphism $\mc FI\to X$.
Moreover, we write ``$\subseteq$'' for the embedding of $I$ into $\mc FI$. Formally this is the embedding given by the unit of the
corresponding monad, which we will spell out later in each concrete case.
The map ``$\subseteq$'' in \cref{H1} is the map acting coordinatewise as the corresponding inclusion.

Provided that $I\times J$ generates $\mc FI\times\mc FJ$, which is the case for convex semilattices, there exists at most one extension 
$\Phi$ and on an abstract level we have an obvious answer to the question (Q).

\begin{itemize}
\item[{\bf(A)}] {\it
	Let $\Lambda\DF\mc F(I\times J)\to\mc FI\times\mc FJ$ be the unique homomorphism extending the inclusion map 
	$I\times J\to\mc FI\times\mc FJ$, and let $\Psi\DF\mc F(I\times J)\to X$ be the unique homomorphism extending $\varphi$.
	Then a homomorphism $\Phi$ with \cref{H1} exists, if and only if\/\footnote{%
		Recall that for any map $f$ its \emph{kernel} is defined as $\ker f=\{(x,y)\DS f(x)=f(y)\}$.%
		}
	\begin{equation}
	\label{H16}
		\ker\Lambda\subseteq\ker\Psi.
	\end{equation}
	}
	\begin{equation}
	\label{H42}
		\begin{tikzcd}[column sep=large]
			& \mc F(I\times J) \arrow[d,"\Lambda",two heads] \arrow[dd,bend left=60,"\Psi"]
			\\
			I\times J \arrow[r,"\subseteq",tail] \arrow[rd,swap,"\varphi"] 
			\arrow[ru,"\subseteq",tail]
			& \mc FI\times\mc FJ \arrow[d,dashed,"\Phi"] 
			\\
			& X
		\end{tikzcd}
	\end{equation}
\end{itemize}
The requirement that $I\times J$ generates $\mc FI\times\mc FJ$ enters here since it is equivalent to surjectivity of $\Lambda$.

Applying this abstract answer requires to determine $\ker\Lambda$ and to be able to check \cref{H16} using only
the initially given map $\varphi$.

In equational classes that contain commutative binary operations there is a simple and explicit condition on $\varphi$ that is always
necessary for a homomorphism $\Phi$ with \cref{H1} to exist.  We refer to this condition as \emph{coordinatewise commutativity}.
An equivalent way to express necessity is to say that a certain relation $\theta$ on 
$\mc FI\times\mc FJ$ generated by the coordinatewise commutativity laws is contained in $\ker\Lambda$. 
See \Cref{H39} where this is elaborated. 

It is not hard to see that in the equational classes of semilattices and of convex algebras coordinatewise commutativity of $\varphi$ 
is always sufficient for $\Phi$ to exist. Equivalently, in these equational classes the relation $\theta$ generates $\ker\Lambda$. 
For the sake of completeness we provide a proof of these facts, cf.\ \Cref{H25} and \Cref{H7}.

Our main focus in the present paper lies on the equational class of convex semilattices. In this class the situation is different: 
we prove in \Cref{H14} that for each two sets $I,J$ having at least two points, the relation $\theta$ does not generate $\ker\Lambda$. 
Hence, for each such $I,J$ there exists a convex semilattice $X$ and a map $\varphi\DF I\times J\to X$ that satisfies coordinatewise
commutativity but does not admit an extension to a homomorphism. 
This is our first main result. It emphasises that convex semilattices have a more complicated structure than their two ingredients 
-- convex algebras and semilattices -- separately. 
Interestingly, it turns out that imposing a fairly weak condition on the range space $X$ of the map $\varphi$ makes coordinatewise
commutativity a sufficient condition for $\Phi$ to exist. We prove this in \Cref{H4} which is our second main result. A possible
intuition is that requiring that $X$ is not too wild compensates for the deficiency that $\theta$ does not generate $\ker\Lambda$. 

Let us explain more about the condition on $X$ which we impose in \Cref{H4}, in particular why we regard it as a weak condition. 
It is based on a property of convex semilattices which we call \emph{midpoint cancellativity}, cf.\ \Cref{H23}. Examples of
midpoint cancellative convex semilattices are the $\min$-$\max$--convex semilattice of any totally ordered set, or the 
$+_p$-$\max$--convex semilattice of any ordered field that contains $\bb R$, cf.\ \Cref{H22}, and there are many more. 
The class of convex semilattices that are admitted in \Cref{H4} is now defined as all isomorphic images of subalgebras of products
of midpoint cancellative convex semilattices. We denote this class as $\bb W$, and it is indeed a large class. For example it contains 
all free convex semilattices, more generally, for any cancellative convex algebra the convex semilattice of its 
finitely generated subalgebras belongs to $\bb W$, cf.\ \Cref{H24}.

The results of the present paper should be put in context with the recent work \cite{woracek:cslsub-arXiv}. In that paper homomorphisms
of free convex semilattices into convex semilattices of a certain class $\mc W$ are studied. This class is
much narrower than the present class $\bb W$: it is defined as all isomorphic images of subalgebras of powers of $\bb R$ (as 
$+_p$-$\max$--convex semilattice). The results from \cite{woracek:cslsub-arXiv} are incomparable with the present ones. 
On the one hand the domain of the
homomorphisms is not restricted to a product of two free ones, but on the other hand the codomain must belong to the
smaller class $\mc W$. However, for $X\in\mc W$ sufficiency of coordinatewise
commutativity could be deduced from \cite{woracek:cslsub-arXiv}.

Finally, let us mention that midpoint cancellativity is a property that is of interest on its own right. 
It will be further studied in our forthcoming work \cite{sokolova.woracek:cslmic-PRE}, where we give a complete description of all
midpoint cancellative convex semilattices. 

\medskip
\noindent
\textbf{Acknowledgement:}
The task to construct homomorphisms whose domain is a product of free convex semilattices
came up in discussions with Stefan Milius and Lutz Schr\"oder during a visit to FAU Erlangen, whom we thank for their hospitality and
fruitful discussions.

\section{The objects of interest}

In this section we start by discussing a crucial property of direct products, which we call coordinatewise commutativity and
which is crucial for the results in this paper. 
Furthermore we recall the definition of the algebras of our present interest: semilattices, convex algebras, and convex semilattices.

\subsection{Coordinatewise commutativity}
\label{H39}

Let $X,Y$ be two sets, each of which is endowed with a binary operation, say $\ast_X$ and $\ast_Y$. 
We endow the direct product $X\times Y$ with the componentwise defined binary operation 
\[
	(x,y)\ast_{X\times Y}(x',y')\DE\big(x\ast_Xx',y\ast_Yy'\big).
\]
Clearly, if $\ast_X$ and $\ast_Y$ are commutative, then $\ast_{X\times Y}$ also is. 
Furthermore, commutativity of $\ast_X$ and $\ast_Y$ implies that $\ast_{X\times Y}$ satisfies the 
following coordinatewise commutativity law:
\begin{equation}
\label{H28}
	\forall x,x'\in X,y,y'\in Y\DP (x,y)\ast_{X\times Y}(x',y')=(x,y')\ast_{X\times Y}(x',y).
\end{equation}
Note that clearly the corresponding law where instead of $y$ and $y'$ the elements $x$ and $x'$ change places also holds.

The law \cref{H28} can be nicely illustrated geometrically. For example consider the set $\bb R$ with the operation 
$+_{\frac 12}\DF (x,y)\mapsto \frac12x+\frac 12y$. Then we can picture \cref{H28} in $\bb R\times\bb R$ as
\begin{center}
\begin{tikzpicture}[x=1.2pt,y=1.2pt,scale=0.8,font=\fontsize{12}{12}]
	\draw[thick,->] (0,0) -- (160,0);
	\draw[thick,->] (0,0) -- (0,100);
	\draw (30,30) -- (130,80);
	\draw (30,80) -- (130,30);
	\draw[dotted] (30,5) -- (30,80);
	\draw[dotted] (130,5) -- (130,80);
	\draw[dotted] (5,30) -- (130,30);
	\draw[dotted] (5,80) -- (130,80);

	\draw[fill,color=green] (30,30) circle [radius=1.5];
	\draw[fill,color=green] (30,80) circle [radius=1.5];
	\draw[fill,color=green] (130,30) circle [radius=1.5];
	\draw[fill,color=green] (130,80) circle [radius=1.5];
	\draw[fill,color=red] (80,55) circle [radius=1.5];

	\draw (30,-10) node {$x$};
	\draw (30,-5) -- (30,5);
	\draw (130,-10) node {$x'$};
	\draw (130,-5) -- (130,5);
	\draw (-10,30) node {$y$};
	\draw (-5,30) -- (5,30);
	\draw (-10,80) node {$y'$};
	\draw (-5,80) -- (5,80);

	\draw[color=red] (210,53) node {$\scriptstyle(x,y)+_{\frac 12}(x',y')=(x,y')+_{\frac 12}(x',y)$};
\end{tikzpicture}
\end{center}
Assume that we have an equational class with a commutative binary operation $\ast$. 
Then the law \cref{H28} implies that the homomorphism $\Lambda$ from {\bf(A)} satisfies
\begin{equation}
\label{H36}
	\forall i,i'\in I,j,j'\in J\DP 
	\Big((i,j)\ast(i',j'),(i,j')\ast(i',j)\Big)\in\ker\Lambda
\end{equation}
In the context of our basic question (Q) we obtain an obvious necessary condition on $\varphi$ for a homomorphism 
$\Phi$ with \cref{H1} to exist. 

\begin{lemma}
\label{H35}
	Assume that in the equational class under consideration there is a commutative binary operation $\ast$. 
	If a homomorphism $\Phi$ as in \cref{H1} exists, then 
	\begin{equation}
	\label{H37}
		\forall i,i'\in I,j,j'\in J\DP 
		\varphi(i,j)\ast\varphi(i',j')=\varphi(i,j')\ast\varphi(i',j)
	\end{equation}
\end{lemma}
\begin{proof} Use \cref{H28} to compute 
	\begin{multline}
	\label{H41}
		\varphi(i,j)\ast\varphi(i',j')=\Phi\big((i,j)\big)\ast\Phi\big((i',j')\big)=
		\Phi\big((i,j)\ast(i',j')\big)
		\\
		=\Phi\big((i,j')\ast(i',j)\big)=\Phi\big((i,j')\big)\ast\Phi\big((i',j)\big)=
		\varphi(i,j')\ast\varphi(i',j).
	\end{multline}
\end{proof}

\noindent
When investigating whether the condition \cref{H37} is also sufficient it is practical to switch between the viewpoints of the 
``extendability property'' as in (Q) on the one hand, and properties of $\ker\Lambda$ as in \cref{H36} on the other. 
We make this precise.

\begin{lemma}
\label{H38}
	Let $\bb V$ be an equational class that contains commutative binary operations $\ast_l$, $l\in L$. 
	Let $I,J$ be nonempty sets, let $\Lambda\DF\mc F(I\times J)\to\mc FI\times\mc FJ$ be the homomorphism extending the
	inclusion map, and set
	\[
		\theta\DE\bigcup_{l\in L}\Big\{\Big((i,j)\ast_l(i',j'),(i,j')\ast_l(i',j)\Big)\DS
		i,i'\in I,j,j'\in J\Big\}.
	\]
	Consider the following statements.
	\begin{Enumerate}
	\item For every algebra $X$ in $\bb V$ and every map $\varphi\DF I\times J\to X$ that satisfies 
		\begin{equation}
		\label{H43}
			\forall l\in L\DQ\forall i,i'\in I,j,j'\in J\DP 
			\varphi(i,j)\ast_l\varphi(i',j')=\varphi(i,j')\ast_l\varphi(i',j)
		\end{equation}
		there exists a homomorphism $\Phi\DF\mc FI\times\mc FJ\to X$ with $\Phi|_{I\times J}=\varphi$.
	\item $\ker\Lambda$ is the smallest congruence containing $\theta$.
	\end{Enumerate}
	Then ${\rm(i)}\Rightarrow{\rm(ii)}$, and if $I\times J$ generates $\mc FI\times\mc FJ$ 
	also ${\rm(ii)}\Rightarrow{\rm(i)}$.
\end{lemma}
\begin{proof}
	Assume (i) holds. Let $\Theta$ be the congruence generated by $\theta$ and denote by
	$\pi\DF\mc F(I\times J)\to\mc F(I\times J)/\raisebox{-2pt}{$\Theta$}$ the canonical projection. We consider the map 
	$\varphi\DE\pi\circ\subseteq\DF I\times J\to\mc F(I\times J)/\raisebox{-2pt}{$\Theta$}$. 
	Since $\theta\subseteq\Theta=\ker\varphi$, the map $\varphi$ satisfies \cref{H43}, and thus $\Phi$ exists. Now we revisit the
	diagram \cref{H42}:
	\[
		\begin{tikzcd}[column sep=large]
			& \mc F(I\times J) \arrow[d,rightarrow,"\Lambda",two heads] \arrow[dd,bend left=60,"\pi"]
			\\
			I\times J \arrow[r,rightarrow,"\subseteq",tail] \arrow[rd,swap,"\varphi=\pi\circ\subseteq"] 
			\arrow[ru,rightarrow,"\subseteq",tail]
			& \mc FI\times\mc FJ \arrow[d,"\Phi"] 
			\\
			& \mc F(I\times J)/\raisebox{-2pt}{$\Theta$}
		\end{tikzcd}
	\]
	By uniqueness of the extension to the free algebra we have $\pi=\Phi\circ\Lambda$, and hence obtain that 
	$\ker\Lambda\subseteq\ker\pi=\Theta$. Since $\theta\subseteq\ker\Lambda$ by \cref{H36}, we also have $\Theta\subseteq\ker\Lambda$ 
	and the statement in (ii) follows. 

	For the proof of ``(ii)$\Rightarrow$(i)'' we simply read \cref{H41} the other way: given $\varphi$ with \cref{H43}, 
	we have 
	\begin{multline*}
		\Psi\big((i,j)\ast_l(i',j')\big)=\Psi\big((i,j)\big)\ast_l\Psi\big((i',j')\big)=
		\varphi(i,j)\ast_l\varphi(i',j')
		\\
		=\varphi(i,j')\ast_l\varphi(i',j)=\Psi\big((i,j')\big)\ast\Psi\big((i',j)\big)=\Psi\big((i,j')\ast(i',j)\big).
	\end{multline*}
	Hence, $\theta\subseteq\ker\Psi$, and by the present assumption (ii) thus also 
	$\ker\Lambda\subseteq\ker\Psi$. Now we can apply (A) and conclude that $\Phi$ exists.
\end{proof}

\subsection{Convex semilattices}
\label{H19}

We recall the definition of convex semilattices. For the sake of completeness we also include semilattices and convex algebras.

\begin{definition}
\label{H33}
	A \emph{semilattice} is a set $X$ together with a binary operation $\oplus$ on $X$ that satisfies 
	\begin{Ilist}
	\item[] \Dis{\forall x\in X\DP x\oplus x=x}
		\hspace*{46mm}(\emph{idempotence})
	\item[] \Dis{\forall x,y\in X\DP x\oplus y=y\oplus x}
		\hspace*{35mm}(\emph{commutativity})
	\item[] \Dis{\forall x,y,z\in X\DP (x\oplus y)\oplus z=x\oplus(y\oplus z)}
		\hspace*{10mm}(\emph{associativity})
	\end{Ilist}
	We write a semilattice as $\langle X,\oplus\rangle$.
\end{definition}

\noindent
A semilattices can equivalently be viewed as a partially ordered set in which each two elements have a supremum.

Next, we recall convex algebras.

\begin{definition}
\label{H34}
	A \emph{convex algebra} is a set $X$ together with a family $\{+_p\DS p\in[0,1]\}$ of binary operations on $X$ that 
	satisfy
	\begin{Ilist}
	\item[] \Dis{\forall x\in X,p\in(0,1)\DP x+_px=x}
		\hspace*{25mm}(\emph{idempotence})
	\item[] \Dis{\forall x,y\in X,p\in(0,1)\DP x+_py=y+_{1-p}x}
		\hspace*{8.5mm}(\emph{parametric commutativity})
	\item[] \Dis{\forall x,y,z\in X,p,q\in(0,1)\DP (x+_py)+_qz=x+_{pq}\Big(y+_{\frac{(1-p)q}{1-pq}}z\Big)}
		\\[1mm] \hspace*{84mm}(\emph{parametric associativity})
	\item[] \Dis{\forall x,y\in X\DP x+_1y=x\wedge x+_0y=y}
		\hspace*{15mm}(\emph{projection axiom})
	\end{Ilist}
	We write a convex algebra as $\langle X,+_p\rangle$.
\end{definition}

\noindent
Let us remark that convex algebras can be axiomatised in different ways; the family of operations can be restricted or expanded.
See e.g.\ \cite{flood:1981,romanowska.smith:1990,sokolova.woracek:pcacon}.

Combining a convex algebra and a semilattice with an appropriate axiom ruling the interplay of these structures leads to the
equational class of convex semilattices.

\begin{definition}
\label{H18}
	A \emph{convex semilattice} is a set $X$ together with binary operations $\{+_p\DS p\in[0,1]\}$ and $\oplus$ on $X$,
	such that 
	\begin{Enumerate}
	\item $\langle X,+_p\rangle$ is a convex algebra,
	\item $\langle X,\oplus\rangle$ is a semilattice,
	\item Each operation $+_p$ \emph{distributes over} $\oplus$, i.e., 
		\[
			\forall x,y,z\in X,p\in[0,1]\DP (x\oplus y)+_pz=(x+_pz)\oplus(y+_pz)
		\]
	\end{Enumerate}
	We write a convex semilattice as $\langle X,+_p,\oplus\rangle$.
\end{definition}

\noindent
The free algebras in the above equational classes can be described explicitly.
Let $I$ be a nonempty set.
\begin{Itemize}
\item The free semilattice $\mc SI$ with basis $I$ is given as
	\begin{align*}
		& \mc SI\DE\big\{A\subseteq I\DS A\text{ finite nonempty}\big\},
		\\
		& A\oplus B\DE A\cup B\quad\text{for }A,B\in\mc SI.
		\mkern265mu
	\end{align*}
	Here the embedding $I\stackrel{\subseteq}{\to}\mc SI$ acts as $i\mapsto\{i\}$, and $\Lambda(\{(i,j)\})=(\{i\},\{j\})$. 
\end{Itemize}
\begin{Itemize}
\item For a subset $M$ of a linear space\footnote{Throughout the paper all linear spaces are understood over the scalar field $\bb R$.}
	(or, more general, of a convex algebra) we denote by $\Conv M$ its \emph{convex hull},
	i.e.\ the smallest convex subset (subalgebra, respectively) containing $M$. 

	Consider the linear space $\bb R^I$, and denote by $e_j\in\bb R^I$ the element $(\delta_{ij})_{i\in I}$ where 
	$\delta_{ij}$ is the Kronecker-delta.  
	The free convex algebra $\mc DI$ with basis $I$ is given as 
	\begin{align*}
		& \mc DI\DE\Conv\{e_i\DS i\in I\}\subseteq\bb R^I,
		\\
		& x+_py\DE px+(1-p)y\quad\text{for }x,y\in\mc DI,p\in[0,1].
		\mkern145mu
	\end{align*}
	Here the embedding $I\stackrel{\subseteq}{\to}\mc DI$ acts as $i\mapsto e_i$, and\footnote{%
		To shorten notation we write $e_{ij}$ for $e_{(i,j)}$.
		}
	$\Lambda(e_{ij})=(e_i,e_j)$. 
\item The free convex semilattice $\mc CI$ with basis $I$ is given as 
	\begin{align*}
		& \mc CI\DE\big\{\Conv M\DSb M\subseteq\mc D,\,M\text{ finite nonempty}\big\},
		\\
		& A+_pB\DE \{px+(1-p)y\DS x\in A,y\in B\}\quad\text{for }A,B\in\mc CI,p\in[0,1],
		\\
		& A\oplus B\DE \Conv(A\cup B)\quad\text{for }A,B\in\mc CI.
	\end{align*}
	Here the embedding $I\stackrel{\subseteq}{\to}\mc CI$ acts as $i\mapsto\{e_i\}$, and $\Lambda(\{e_{ij}\})=(\{e_i\},\{e_j\})$. 
\end{Itemize}
Concerning the categorical background of these algebras we mention several facts: 
Semilattices are the Eilenberg-Moore algebras of the nonempty finite powerset monad. Convex algebras are the Eilenberg-Moore algebras
of the distribution monad $\mc D$, see \cite{doberkat:2006,doberkat:2008,jacobs.westerbaan.westerbaan:2015}.
Convex semilattices are the Eilenberg-Moore algebras of the monad $\mc C$
of nonempty finitely generated convex subsets of distributions, see \cite{jacobs:2008}
with origins in \cite{mislove:2000,varacca:2003,varacca.winskel:2006,goubaultlarrecq:2008,tix.keimel.plotkin:2009} for $\mc C$ being a
monad, and \cite[\S3.2]{bonchi.sokolova.vignudelli:2022,bonchi.sokolova.vignudelli:2021} for the description of its
Eilenberg-Moore algebras.
Finally, we mention that the construction leading from $\mc DI$ to $\mc CI$ is functorial and this fact will be used later. 

\begin{remark}
\label{H29}
	Let $\langle X,+_p\rangle$ be a convex algebra. The set of all nonempty finitely generated subalgebras of $X$ can be endowed
	with operations $+_p,\oplus$ in a natural way:
	\begin{align*}
		& \Moewe X\DE\big\{\Conv M\DSb M\subseteq X,\,M\text{ finite nonempty}\big\},
		\\
		& A+_pB\DE \{x+_py\DS x\in A,y\in B\}\quad\text{for }A,B\in\Moewe X,p\in[0,1],
		\\
		& A\oplus B\DE \Conv(A\cup B)\quad\text{for }A,B\in\Moewe X.
	\end{align*}
	More explicitly,
	\[
		A\oplus B=\big\{a+_pb\DS a\in A,b\in B,p\in[0,1]\big\}=\bigcup_{p\in[0,1]}(A+_pB).
	\]
	Then $\langle\Moewe X,+_p,\oplus\rangle$ is a convex semilattice. 
	Note the difference with the functor $\mc P_c$ from \cite[\S5.1]{bonchi.silva.sokolova:2021}: we consider only the finitely
	generated nonempty convex subsets, while $\mc P_c$ takes all nonempty convex subsets.

	Assume we have convex algebras $\langle X,+_p\rangle$ and $\langle Y,+_p\rangle$, and a homomorphism 
	$\varphi\DF X\to Y$ of convex algebras. We define $\Moewe[12][8][7]\varphi$ as the image map
	\[
		\Moewe[12][8][7]\varphi\DF\left\{
		\begin{array}{rcl}
			\Moewe X & \to & \Moewe Y
			\\
			A & \mapsto & \{\varphi(a)\DS a\in A\}
		\end{array}
		\right.
	\]
	Then $\Moewe[12][8][7]\varphi$ is a homomorphism of convex semilattices. 
	The assignment defined in that way satisfies
	\[
		\Moewe[22][1][10]{\Id_X}=\Id\mkern-2mu\raisebox{-4pt}{\scalebox{0.7}{$\Moewe X$}},\qquad
		\Moewe[12][8][7]\varphi\circ\Moewe[12][8][10]\psi=\Moewe[35][-18][12]{(\varphi\circ\psi)}.
	\]
\end{remark}

\noindent
We return to the discussion of our question (Q). Each of the above equational classes contains a commutative binary operation: 
$\oplus$ or $+_{\frac 12}$ or both, respectively. \Cref{H35} instantiates as follows.

\begin{corollary}
\label{H26}
	Let $X$ be a set, let $I,J$ be nonempty sets, and $\varphi\DF I\times J\to X$.
	\begin{Enumerate}
	\item Assume that $X$ carries a semilattice structure. 
		If there exists a homomorphism $\Phi\DF\mc SI\times\mc SJ\to X$ with \cref{H1}, then 
		\begin{equation}
		\label{H3}
			\forall i,i'\in I,j,j'\in J\DP 
			\varphi(i,j)\oplus\varphi(i',j')=\varphi(i,j')\oplus\varphi(i',j)
		\end{equation}
	\item Assume that $X$ carries a convex algebra structure. 
		If there exists a homomorphism $\Phi\DF\mc DI\times\mc DJ\to X$ with \cref{H1}, then 
		\begin{equation}
		\label{H2}
			\forall i,i'\in I,j,j'\in J\DP 
			\varphi(i,j)+_{\frac 12}\varphi(i',j')=\varphi(i,j')+_{\frac 12}\varphi(i',j)
		\end{equation}
	\item Assume that $X$ carries a convex semilattice structure. 
		If there exists a homomorphism $\Phi\DF\mc CI\times\mc CJ\to X$ with \cref{H1}, then 
		\cref{H3} and \cref{H2} both hold.
	\end{Enumerate}
\end{corollary}

\section{Sufficiency for convex algebras and semilattices}

For convex algebras and semilattices the condition arising from coordinatewise commutativity 
turns out to be also sufficient. 

We first settle the case of semilattices, where it is very easy to directly show existence of $\Phi$. 

\begin{proposition}
\label{H25}
	Let $\langle X,\oplus\rangle$ be a semilattice, let $I,J$ be nonempty sets, and $\varphi\DF I\times J\to X$. 
	If $\varphi$ satisfies \cref{H3}, then there exists a homomorphism $\Phi\DF\mc SI\times\mc SJ\to X$ with 
	$\Phi|_{I\times J}=\varphi$.
\end{proposition}
\begin{proof}
	Define a map $\Phi\DF\mc SI\times\mc SJ\to X$ as 
	\[
		\Phi((A,B))\DE\bigoplus_{a\in A}\bigoplus_{b\in B}\varphi(a,b)
		\quad\text{for }A\in\mc SI,B\in\mc SJ.
	\]
	Note here that $A$ and $B$ are finite. We check that $\Phi$ is a homomorphism of semilattices. 
	Let $A,C\in\mc SI$, $B,D\in\mc SJ$, and compute 
	\begin{align*}
		& \Phi((A,B))\oplus\Phi((C,D))
		=\Big[\bigoplus_{a\in A}\bigoplus_{b\in B}\varphi(a,b)\Big]\oplus
		\Big[\bigoplus_{c\in C}\bigoplus_{d\in D}\varphi(c,d)\Big]
		\\
		& =
		\bigoplus_{a\in A}\bigoplus_{b\in B}\bigoplus_{c\in C}\bigoplus_{d\in D}\big[\varphi(a,b)\oplus\varphi(c,d)\big]
		\\
		& \stackrel{\cref{H3}}{=}
		\bigoplus_{a\in A}\bigoplus_{b\in B}\bigoplus_{c\in C}\bigoplus_{d\in D}
		\Big(\big[\varphi(a,b)\oplus\varphi(c,d)\big]\oplus\big[\varphi(a,d)\oplus\varphi(c,b)\big]\Big)
		\\
		& =\Big[\bigoplus_{a\in A}\bigoplus_{b\in B}\varphi(a,b)\Big]\!\!\oplus\!\!
		\Big[\bigoplus_{c\in C}\bigoplus_{d\in D}\varphi(c,d)\Big]\!\!\oplus\!\!
		\Big[\bigoplus_{a\in A}\bigoplus_{d\in D}\varphi(a,d)\Big]\!\!\oplus\!\!
		\Big[\bigoplus_{c\in C}\bigoplus_{b\in B}\varphi(c,b)\Big]
		\\
		& =\Big[\bigoplus_{x\in A\cup C}\bigoplus_{y\in B\cup D}\varphi(x,y)\Big]
		=\Phi\big((A,B)\oplus(C,D)\big).
	\end{align*}
\end{proof}

\noindent
We turn to convex algebras. There the result is less obvious and we proceed via $\ker\Lambda$.

\begin{proposition}
\label{H7}
	Let $I,J$ be nonempty sets, and let $\Theta$ be the congruence on $\mc D(I\times J)$ generated by 
	\[
		\theta\DE\Big\{\Big((e_i,e_j)+_{\frac 12}(e_{i'},e_{j'}),(e_i,e_{j'})+_{\frac 12}(e_{i'},e_j)\Big)\DSB
		i,i'\in I,j,j'\in J\Big\}.
	\]
	Then $\ker\Lambda=\Theta$.
\end{proposition}

\noindent
The proof uses an inductive procedure. To formulate it, we introduce some notation. 
Let $x=\sum\limits_{i\in I,j\in J}p_{ij}(x)e_{ij}\in\mc D(I\times J)$. 
We denote the row sums and column sums of $p_{ij}(x)$ as
\[
	p^{(1)}_i(x)\DE\sum_{j\in J}p_{ij}(x),\quad p^{(2)}_j(x)\DE\sum_{i\in I}p_{ij}(x).
\]
Note that $x$ is a coupling of $\sum_{i\in I}p^{(1)}_ie_i$ and $\sum_{j\in J}p^{(2)}_je_j$.
Moreover, we set 
\begin{align*}
	& I_x\DE\big\{i\in I\DS p_i^{(1)}(x)>0\big\},\quad J_x\DE\big\{j\in J\DS p_j^{(2)}(x)>0\big\},
	\\
	& M_x\DE\big\{(i,j)\in I\times J\DS p_{ij}(x)\geq p^{(1)}_i(x)p^{(2)}_j(x)\big\}.
\end{align*}
If, in addition, $(k,l)\in I\times J$ is given, then we set 
\[
	M_x^{kl}\DE\big\{(k,j)\in I\times J\DS p_{kj}(x)>p^{(1)}_k(x)p^{(2)}_j(x)\big\}
	\cup\big\{(i,l)\in I\times J\DS p_{il}(x)>p^{(1)}_i(x)p^{(2)}_l(x)\big\}.
\]
Since $x\in\mc D(I\times J)$ is a finite convex combination of the vectors $e_{ij}$, both sets 
$I_x$ and $J_x$ are finite. Also, observe that 
\[
	(I\times J)\setminus(I_x\times J_x)\subseteq M_x,\quad M_x^{kl}\subseteq I_x\times J_x.
\]
The idea of the inductive argument is that we can pass to a $\Theta$-equivalent of $x$ that has the same marginals as $x$ and make 
either $M_x$ increase or remain the same while some $M_x^{kl}$ decreases.
We formulate this as a separate lemma. 

\begin{lemma}
\label{H31}
	Assume $x=\sum\limits_{i\in I,j\in J}p_{ij}e_{ij}\in\mc D(I\times J)$ and $(k,l)\in(I\times J)\setminus M_x$, in particular, 
	$(k,l)\in I_x\times J_x$. Then there exists $y\in\mc D(I\times J)$ such that 
	\begin{Enumerate}
	\item $x\,\Theta\,y$,
	\item $\forall i\in I\DP p^{(1)}_i(x)=p^{(1)}_i(y)$ and $\forall j\in J\DP p^{(2)}_j(x)=p^{(2)}_j(y)$
	\item $M_y\supseteq M_x\cup\{(k,l)\}$ or $M_y=M_x\wedge M_y^{kl}\subset M_x^{kl}$.
	\end{Enumerate}
\end{lemma}
\begin{proof}
	To shorten notation, we drop the argument $x$ from $p_{ij},p^{(1)}_i,p^{(2)}_j$. 

	Since $(k,l)\notin M_x$ we have $p_{kl}<p^{(1)}_kp^{(2)}_l$. 
	Since 
	\[
		\sum_{i\in I}p_{il}=p^{(2)}_l=\sum_{i\in I}p^{(1)}_ip^{(2)}_l
		\quad\text{and}\quad
		\sum_{j\in J}p_{kj}=p^{(1)}_k=\sum_{j\in J}p^{(1)}_kp^{(2)}_j
	\]
	there exist $k'\in I$ and $l'\in J$ such that $p_{k'l}>p^{(1)}_{k'}p^{(2)}_l$ and $p_{kl'}>p^{(1)}_kp^{(2)}_{l'}$. 
	Clearly, 
	\[
		k'\neq k,\ l'\neq l,\quad (k,l'),(k',l)\in M_x^{kl},\quad k,k'\in I_x,\ l,l'\in J_x.
	\]
	Set 
	\[
		s\DE\min\big\{p_{k'l}-p^{(1)}_{k'}p^{(2)}_l,p_{kl'}-p^{(1)}_kp^{(2)}_{l'}\big\},
	\]
	then 
	\[
		0<2s\leq(p_{k'l}-p^{(1)}_{k'}p^{(2)}_l)+(p_{kl'}-p^{(1)}_kp^{(2)}_{l'})\leq p_{k'l}+p_{kl'}\leq 1.
	\]
	\begin{Ilist}
	\item Consider first the case that $2s=1$. Since $s\leq p_{k'l}$ and $s\leq p_{kl'}$, it follows that 
		$p_{k'l}=p_{kl'}=\frac 12$ and in turn that $p_{ij}=0$ for all $(i,j)\notin\{(k',l),(k,l')\}$. We see that 
		$x=\frac 12e_{k'l}+\frac 12e_{kl'}$. Therefore $x\,\theta\,(\frac 12e_{kl}+\frac 12e_{k'l'})$, 
		and hence
		\[
			x=\tfrac 12x+\tfrac 12x\ \Theta\ 
			\tfrac 12\big(\tfrac 12e_{k'l}+\tfrac 12e_{kl'}\big)+
			\tfrac 12\big(\tfrac 12e_{kl}+\tfrac 12e_{k'l'}\big)
			=\tfrac 14e_{kl}+\tfrac 14e_{k'l}+\tfrac 14e_{kl'}+\tfrac 14e_{k'l'}\ED y.
		\]
		Property (i) holds by definition, and (ii) holds since $I_x=\{k,k'\}=I_y$ and $J_x=\{l,l'\}=J_y$. 
		Finally, we have $M_y=I\times J$ and hence (iii) holds. 
	\item Consider now the case that $2s<1$. Set
		\begin{align*}
			p_{ij}'\DE &\, 
			\begin{cases}
				p_{ij}-s \CAS (i,j)\in\{(k',l),(k,l')\},
				\\
				p_{ij} \CASO,
			\end{cases}
			\\
			p_{ij}''\DE &\, 
			\begin{cases}
				p_{ij}+s \CAS (i,j)\in\{(k,l),(k',l')\},
				\\
				p_{ij}-s \CAS (i,j)\in\{(k',l),(k,l')\},
				\\
				p_{ij} \CASO,
			\end{cases}
		\end{align*}
		and observe that $\sum\limits_{i\in I,j\in J}p_{ij}''e_{ij}\in\mc D(I\times J)$, 
		and that the row sums and column sums of $p_{ij}$ and $p_{ij}''$ coincide.

		We compute 
		\begin{align*}
			x= &\, \sum_{i\in I,j\in J}p_{ij}e_{ij}
			=(1-2s)\Big[\sum_{i\in I,j\in J}\tfrac{p_{ij}'}{1-2s}e_{ij}\Big]
			+2s\big(\tfrac 12e_{k'l}+\tfrac 12e_{kl'}\big)
			\\
			\Theta &\, \ 
			(1-2s)\Big[\sum_{i\in I,j\in J}\tfrac{p_{ij}'}{1-2s}e_{ij}\Big]
			+2s\big(\tfrac 12e_{kl}+\tfrac 12e_{k'l'}\big)
			=\sum_{i\in I,j\in J}p_{ij}''e_{ij}\ED y.
		\end{align*}
		Property (i) holds by definition, and (ii) holds since row and column sums of $x$ and $y$ are the same.

		Next, we show that $M_y\supseteq M_x\cup\{(k',l),(k,l')\}$.
		If $(i,j)\notin\{k,k'\}\times\{l,l'\}$ we have $p_{ij}''=p_{ij}$, and hence 
		for such indices $(i,j)\in M_x$ if and only if $(i,j)\in M_y$. 
		Moreover, we have 
		\[
			p_{k'l}''=p_{k'l}-s\geq p^{(1)}_{k'}p^{(2)}_l,\ p_{kl'}''=p_{kl'}-s\geq p^{(1)}_kp^{(2)}_{l'},\quad
			p_{kl}''>p_{kl},\ p_{k'l'}''>p_{k'l'}.
		\]
		This shows that $(k',l),(k,l')\in M_y$, that $(k,l)\in M_x$ implies $(k,l)\in M_y$, and that
		$(k',l')\in M_x$ implies $(k',l')\in M_y$.

		If $(k,l)\in M_y$, we are done. Assume that $(k,l)\notin M_y$. 
		If $(i,j)\notin\{(k,l),(k,l'),(k',l)\}$, then $(i,j)\in M_x^{kl}$ if and only if $(i,j)\in M_y^{kl}$. 
		We have $(k,l)\notin M_x$ and $(k,l)\notin M_y$, and hence also $(k,l)\notin M_x^{kl}$ and $(k,l)\notin M_y^{kl}$.
		By our choice of $s$ at least one of $p_{k'l}''=p^{(1)}_{k'}p^{(2)}_l$ and 
		$p_{kl'}''=p^{(1)}_kp^{(2)}_{l'}$ holds. 
		Hence, at least one of $(k,l')$ and $(k',l)$ does not belong to $M_y^{kl}$. Both of these indices 
		belong to $M_x^{kl}$, and we see that $M_y^{kl}\subset M_x^{kl}$. 
	\end{Ilist}
\end{proof}

\begin{corollary}
\label{H44}
	For all $x=\sum\limits_{i\in I,j\in J}p_{ij}e_{ij}\in\mc D(I\times J)$ it holds that 
	\[
		x\mkern9mu \Theta\sum_{i\in I,j\in J}p^{(1)}_ip^{(2)}_je_{ij}.
	\]
\end{corollary}
\begin{proof}
	We proceed by induction on $|(I\times J)\setminus M_x|$, which is a natural number for any $x\in\mc D(I\times J)$. 
	\begin{Ilist}
	\item If $M_x=I\times J$, then $p_{ij}\geq p^{(1)}_ip^{(2)}_j$ for all $i\in I$, $j\in J$. Since the total sums of both sides 
		are equal to $1$, it follows that $x=\sum\limits_{i\in I,j\in J}p^{(1)}_ip^{(2)}_je_{ij}$ and we are done. 
	\item Assume we have $x$ with $M_x\neq I\times J$. Choose $(k,l)\notin M_x$. The set $M_x^{kl}$ is finite. Hence, after
		finitely many applications of \Cref{H31} we obtain an element $y\in\mc D(I\times J)$ with $x\Theta y$, 
		$M_y\supset M_x$, and which has the same row and column sums as $x$.
	\end{Ilist}
\end{proof}

\begin{proof}[Proof of \Cref{H7}]
	To show that $\ker\Lambda=\Theta$, it is enough to note that $\Lambda$ acts as 
	\[
		\Lambda\Big(\sum_{i\in I,j\in J}p_{ij}e_{ij}\Big)=
		\Big(\sum_{i\in I}p^{(1)}_ie_i,\sum_{j\in J}p^{(2)}_je_j\Big),
	\]
	and refer to \Cref{H44}.
\end{proof}

\section{Non-sufficiency for convex semilattices}

In the equational class of convex semilattices we have the two commutative operations $+_{\frac 12}$ and $\oplus$. The relation 
$\theta$ from \Cref{H38} thus is
\begin{equation}
\label{H6}
	\begin{aligned}
		\theta\DE &\, 
		\Big\{
		\big(\{e_{ij}\}+_{\frac 12}\{e_{i'j'}\},\{e_{ij'}\}+_{\frac 12}\{e_{i'j}\}\big)
		\DSB i,i'\in I,j,j'\in J\Big\}
		\\[2mm]
		&\, \cup\Big\{
		\big(\{e_{ij}\}\oplus\{e_{i'j'}\},\{e_{ij'}\}\oplus\{e_{i'j}\}\big)
		\DSB i,i'\in I,j,j'\in J\Big\}.
	\end{aligned}
\end{equation}
The next theorem is our first main result. 

\begin{theorem}
\label{H14}
	Let $I,J$ be two sets with $|I|,|J|\geq 2$, let $\Lambda\DF\mc C(I\times J)\to\mc CI\times\mc CJ$ be the homomorphism 
	with 
	\[
		\forall(i,j)\in I\times J\DP \Lambda(\{e_{ij}\})=\big(\{e_i\},\{e_j\}\big),
	\]
	and let $\Theta$ be the smallest congruence relation on $\mc C(I\times J)$ that contains \cref{H6}. Then
	\[
		\Theta\neq\ker\Lambda.
	\]
\end{theorem}

\noindent
The strategy for the proof of \Cref{H14} is to figure out a property of relations that $\Theta$ has, 
but $\ker\Lambda$ does not have.

Recall the functor $\Moewe[12][8][7]{\phantom{\varphi}}$ from \Cref{H29}.

\begin{definition}
\label{H10}
	Let $V$ be a linear space and $W\subseteq V$ a linear subspace. 
	We say that a relation $\Gamma\subseteq\Moewe V\times\Moewe V$ \emph{has the property} $(W)$, if 
	\[
		\forall (A,B)\in\Gamma,w\in V\DP A\subseteq w+W\Rightarrow B\subseteq w+W
	\]
\end{definition}

\noindent
For later reference we state a couple of obvious properties of this notion. 

\begin{lemma}
\label{H11}
	Let $V$ be a linear space and $W\subseteq V$ a linear subspace. 
	\begin{Enumerate}
	\item The diagonal $\Delta\subseteq\Moewe V\times\Moewe V$ has $(W)$.
	\item If $\Gamma,\Gamma'\subseteq\Moewe V\times\Moewe V$ with $\Gamma\subseteq\Gamma'$ and $\Gamma'$ has $(W)$, 
		then $\Gamma$ has $(W)$.
	\item If $\Gamma_i\subseteq\Moewe V\times\Moewe V$ for $i\in I$ and all $\Gamma_i$ have $(W)$, then 
		$\bigcup_{i\in I}\Gamma_i$ has $(W)$.
	\item If $\Gamma\subseteq\Moewe V\times\Moewe V$ has $(W)$, then the transitive closure of $\Gamma$ has 
		$(W)$.
	\end{Enumerate}
\end{lemma}

\noindent
Next we consider the operators needed to build a congruence:
for a relation $\Gamma\subseteq\Moewe V\times\Moewe V$ denote, 
\begin{align*}
	\tau(\Gamma)\DE &\, \big\{(A+_pA',B+_pB')\DS (A,B),(A',B')\in\Gamma,p\in[0,1]\big\},
	\\
	\sigma(\Gamma)\DE &\, \big\{(A\oplus A',B\oplus B')\DS (A,B),(A',B')\in\Gamma\big\}.
\end{align*}
We have $\Gamma\subseteq\tau(\Gamma)\cap\sigma(\Gamma)$. Moreover, if $\Gamma$ is symmetric, then
$\tau(\Gamma)$ and $\sigma(\Gamma)$ also are symmetric. 

\begin{lemma}
\label{H12}
	If $\Gamma\subseteq\Moewe V\times\Moewe V$ has $(W)$, then $\tau(\Gamma)$ and $\sigma(\Gamma)$ 
	have $(W)$.
\end{lemma}
\begin{proof}
	We first consider $\sigma(\Gamma)$. Let $(A,B),(A',B')\in\Gamma$ and assume that $A\oplus A'\subseteq w+W$. 
	Since $A\subseteq A\oplus A'$, we have $A\subseteq w+W$ and consequently $B\subseteq w+W$. It follows in the same way
	that $B'\subseteq w+W$. The set $w+W$ is convex, and therefore $B\oplus B'\subseteq w+W$. 

	Now consider $\tau(\Gamma)$. Let $(A,B),(A',B')\in\Gamma$, $p\in[0,1]$, and assume that $A+_pA'\subseteq w+W$. 
	If $p=1$ this means that $A\subseteq w+W$ and we obtain $B+_pB'=B\subseteq w+W$. The case that $p=0$ is analogous.
	Assume that $p\in(0,1)$. First, we show that 
	\begin{equation}
	\label{H15}
		\forall x,y\in A\DP x-y\in W,\qquad \forall x',y'\in A'\DP x'-y'\in W.
	\end{equation}
	Assume we have $x,y\in A$. Fix $a'\in A'$. Then $x+_pa',y+_pa'\in A+_pA'\subseteq w+W$, and hence
	\[
		p(x-y)=[px+(1-p)a']-[py+(1-p)a']=(x+_pa')-(y+_pa')\in W.
	\]
	Since $p\neq 0$, it follows that $x-y\in W$. The statement for $A'$ follows in the same way using that $p\neq 1$.

	Now fix $x\in A$ and $x'\in A'$. By \cref{H15} we have $A\subseteq x+W$ and $A'\subseteq x'+W$. It follows that 
	$B\subseteq x+W$ and $B'\subseteq x'+W$, and thus 
	\[
		B+_pB'\subseteq(px+(1-p)x')+W\subseteq (A+_pA')+W\subseteq(w+W)+W=w+W.
	\]
\end{proof}

\begin{corollary}
\label{H13}
	Assume $\gamma\subseteq\Moewe V\times\Moewe V$ is symmetric, and let $\Gamma\subseteq\Moewe V\times\Moewe V$ 
	be the smallest congruence relation that contains $\gamma$. If $\gamma$ has $(W)$, then $\Gamma$ has $(W)$. 
\end{corollary}
\begin{proof}
	Denote again by $\Delta$ the diagonal in $\Moewe V\times\Moewe V$. Since $\gamma$ is symmetric, $\Gamma$ is the 
	transitive closure of 
	\[
		(\Delta\cup\gamma)\cup\tau(\Delta\cup\gamma)\cup[\sigma\circ\tau](\Delta\cup\gamma)
		\cup[\tau\circ\sigma\circ\tau](\Delta\cup\gamma)\cup\ldots
	\]
	Now apply \Cref{H11} and \Cref{H12}.
\end{proof}

\noindent
We have collected everything that is needed for the proof of \Cref{H14}.

\begin{proof}[Proof of \Cref{H14}]
	Let $V$ be the linear space 
	\[
		V\DE\Span\big\{e_{ij}\DS i\in I,j\in J\big\}\subseteq\bb R^{I\times J},
	\]
	and let $\lambda\DF V\to\bb R^I\times\bb R^J$ be the linear map with 
	\[
		\forall i\in I,j\in J\DP \lambda(e_{ij})=(e_i,e_j)
	\]
	Note here that $\{e_{ij}\DS i\in I,j\in J\}$ is a basis of $V$. Moreover, note that $\mc D(I\times J)\subseteq V$ and
	hence $\mc C(I\times J)\subseteq\Moewe V$. 

	Throughout the following fix $i,i'\in I$ and $j,j'\in J$ with $i\neq i'$ and $j\neq j'$, 
	and let $W\subseteq V$ be the linear subspace 
	\[
		W\DE\Span\Big(\lambda^{-1}(\{0\})\cup
		\big\{\tfrac 12e_{ij}+\tfrac 12e_{i'j}-e_{ij'}\big\}\Big).
	\]
	Moreover, let $\theta$ be the relation \cref{H6} and let $\Theta$ be the congruence generated by $\theta$.
	We are going to show that $\ker\Lambda\subseteq\Moewe V\times\Moewe V$ does not have the property $(W)$, while 
	$\Theta$ has. 
	\begin{Ilist}
	\item 
	In the first step we deal with $\ker\Lambda$. Consider the sets 
	\[
		A\DE\Big[\{e_{ij}\}+_{\frac 12}\{e_{i'j}\}\Big]\oplus\{e_{ij'}\},\quad
		B\DE\Big[\{e_{ij'}\}+_{\frac 12}\{e_{i'j'}\}\Big]\oplus\{e_{ij}\}.
	\]
	Then 
	\begin{align*}
		\Lambda(A)= &\, 
		\Lambda\Big(\Big[\{e_{ij}\}+_{\frac 12}\{e_{i'j}\}\Big]\oplus\{e_{ij'}\}\Big)
		=\Big[\big(\{e_i\},\{e_j\}\big)+_{\frac 12}\big(\{e_{i'}\},\{e_j\}\big)\Big]\oplus
		\big(\{e_i\},\{e_{j'}\}\big)
		\\
		= &\, \big(\{e_i\}+_{\frac 12}\{e_{i'}\},\{e_j\}\big)\oplus
		\big(\{e_i\},\{e_{j'}\}\big)
		=\Big(\big[\{e_i\}+_{\frac 12}\{e_{i'}\}\big]\oplus\{e_i\},\{e_j\}\oplus\{e_{j'}\}\Big)
		\\
		= &\, \Big(\big[\{e_i\}+_{\frac 12}\{e_{i'}\}\big]\oplus\{e_i\},\{e_{j'}\}\oplus\{e_j\}\Big)
		=\big(\{e_i\}+_{\frac 12}\{e_{i'}\},\{e_{j'}\}\big)\oplus
		\big(\{e_i\},\{e_j\}\big)
		\\
		= &\, \Big[\big(\{e_i\},\{e_{j'}\}\big)+_{\frac 12}\big(\{e_{i'}\},\{e_{j'}\}\big)\Big]\oplus
		\big(\{e_i\},\{e_j\}\big)
		=\Lambda\Big(\Big[\{e_{ij'}\}+_{\frac 12}\{e_{i'j'}\}\Big]\oplus\{e_{ij}\}\Big)
		\\
		= &\, \Lambda(B),
	\end{align*}
	i.e., $(A,B)\in\ker\Lambda$.

	Using that $0,\tfrac 12e_{ij}+\tfrac 12e_{i'j}-e_{ij'}\in W$, we obtain $A\subseteq e_{ij'}+W$. 
	Assume towards a contradiction that $B\subseteq e_{ij'}+W$. 
	Then $[e_{ij'}+_{\frac 12}e_{i'j'}]-e_{ij}\in B-B\subseteq W$. Choose $\alpha\in\bb R$ such that 
	\[
		\big[\tfrac 12e_{ij'}+\tfrac 12e_{i'j'}-e_{ij}\big]+
		\alpha\big[\tfrac 12e_{ij}+\tfrac 12e_{i'j}-e_{ij'}\big]\in\lambda^{-1}(\{0\}).
	\]
	Applying $\lambda$ gives 
	\[
		\Big((1+\alpha)\big(-\tfrac 12e_i+\tfrac 12e_{i'}\big),
		(1-\alpha)\big(e_{j'}-e_j\big)\Big)=0.
	\]
	Since $i\neq i'$ and $j\neq j'$ it follows that $1+\alpha=0$ and $1-\alpha=0$. We have reached a contradiction.
	\item
	In the second step we show that $\theta$ has the property $(W)$, which implies by \Cref{H13} that $\Theta$ has $(W)$. 
	Let $l,l'\in I$ and $k,k'\in J$. If $l=l'$ or $k=k'$ the required implication is trivial, hence assume that 
	$l\neq l'$ and $k\neq k'$. First, we compute 
	\[
		\lambda\big(\tfrac 12e_{lk}+\tfrac 12e_{l'k'}\big)= 
		\tfrac 12(e_l,e_k)+\tfrac 12(e_{l'},e_{k'})
		=\tfrac 12(e_l,e_{k'})+\tfrac 12(e_{l'},e_k)=
		\lambda\big(\tfrac 12e_{lk'}+\tfrac 12e_{l'k}\big).
	\]
	This implies that
	\[
		\big(\tfrac 12e_{lk}+\tfrac 12e_{l'k'}\big)-\big(\tfrac 12e_{lk'}+\tfrac 12e_{l'k}\big)
		\in \lambda^{-1}(\{0\})\subseteq W.
	\]
	and hence 
	\[
		\forall w\in V\DP \Big(
		\{e_{lk}\}+_{\frac 12}\{e_{l'k'}\}\subseteq w+W\ \Leftrightarrow\ 
		\{e_{lk'}\}+_{\frac 12}\{e_{l'k}\}\subseteq w+W\Big)
	\]
	Next, we show that 
	\[
		\forall w\in V\DP \{e_{lk}\}\oplus\{e_{l'k'}\}\nsubseteq w+W
	\]
	Assume towards a contradiction that we have $w\in V$ such that both, $e_{lk}$ and $e_{l'k'}$ belong to $w+W$.
	Then $e_{lk}-e_{l'k'}\in W$. Choose $\alpha\in\bb R$ such that 
	\[
		\big[e_{lk}-e_{l'k'}\big]+
		\alpha\big[\tfrac 12e_{ij}+\tfrac 12e_{i'j}-e_{ij'}\big]\in\lambda^{-1}(\{0\}).
	\]
	Applying $\lambda$ gives 
	\[
		\Big(e_l-e_{l'}+\alpha\big(-\tfrac 12e_i+\tfrac 12e_{i'}\big),
		e_k-e_{k'}+\alpha\big(e_j-e_{j'}\big)\Big)=0.
	\]
	The first component implies that either $l=i,l'=i',\alpha=2$ or $l=i',l'=i,\alpha=-2$. Hence, $\alpha\in\{-2,2\}$. 
	At the same time, the second component implies that either $k=j,k'=j',\alpha=-1$ or $k=j',k'=j,\alpha=1$, so 
	$\alpha\in\{-1,1\}$, a contradiction.
	\end{Ilist}
\end{proof}

\section{Sufficiency for a class of convex semilattices}

For convex semilattices $\langle X,+_p,\oplus\rangle$ possessing a certain additional property, the conditions 
\cref{H3} and \cref{H2} are sufficient in order that $\Phi$ exists (despite that in general they are not). 
This is our second main result and is given in \Cref{H4}.

First we digress to explain the mentioned additional property.

\subsection{Midpoint cancellative convex semilattices}

The following cancellation property is closely related with the 
coordinatewise commutativity law \cref{H28} that holds for $+_{\frac 12}$ and $\oplus$ in a product of convex semilattices.

\begin{definition}
\label{H23}
	Let $\langle X,+_p,\oplus\rangle$ be a convex semilattice. We say that $X$ is 
	\emph{midpoint cancellative}, if 
	\[
		\forall x,y,z,w\in X\DP 
		x+_{\frac 12}y=z+_{\frac 12}w\wedge x\oplus y=z\oplus w
		\Rightarrow (x,y)=(z,w)\vee(x,y)=(w,z)
	\]
\end{definition}

\noindent
This cancellation property is obviously inherited by subalgebras. Contrasting this, for direct products a negative 
statement holds: if $X$ and $Y$ are convex semilattices 
with $|X|,|Y|\geq 2$, then their product $X\times Y$ is not midpoint cancellative. 
The reason is coordinatewise commutativity: let $x,x'\in X$, $y,y'\in Y$ with $x\neq x'$, $y\neq y'$, then 
\[
	(x,y)+_{\frac 12}(x',y')=(x,y')+_{\frac 12}(x',y)\wedge(x,y)\oplus(x',y')=(x,y')\oplus(x',y),
\]
but 
\[
	\big((x,y),(x',y')\big)\neq\big((x,y'),(x',y)\big)\wedge
	\big((x,y),(x',y')\big)\neq\big((x',y),(x,y')\big).
\]
Let us give two examples of midpoint cancellative convex semilattices. 

\begin{example}
\label{H22}
	\phantom{}
	\begin{Enumerate}
	\item Let $X$ be a totally ordered set. For $x,y\in X$ set
		\[
			x\oplus y\DE \max\{x,y\},\qquad
			x+_py\DE 
			\begin{cases}
				y\CAS p=0,
				\\
				\min\{x,y\}\CAS p\in(0,1),
				\\
				x\CAS p=1.
			\end{cases}
		\]
		Then $\langle X,+_p,\oplus\rangle$ is a midpoint cancellative convex semilattice. 
	\item Let $K$ be an ordered field that contains $\bb R$. Set 
		\[
			x\oplus y\DE\max\{x,y\},\ x+_py\DE px+(1-p)y\quad\text{for } x,y\in K,p\in[0,1]
			.
		\]
		Then $\langle K,+_p,\oplus\rangle$ is a midpoint cancellative convex semilattice. 

		This fact applies in particular with the ordered field $\bb R$ itself. 
		Whenever we write ``$\bb R$'' for a convex semilattice we mean the midpoint cancellative convex semilattice 
		$\langle\bb R,+_p,\oplus\rangle$.
	\end{Enumerate}
\end{example}

\noindent
We point out that in these two examples the convex algebra structure behaves very differently: 
for each two $x\neq y$ the map $p\mapsto x+_py$ is 
in (i) constant for $p\in(0,1]$ if $x<y$ (for $p\in[0,1)$ if $x>y$, respectively), while 
in (ii) it is injective on $[0,1]$. 

\begin{remark}
\label{H20}
	It is not a coincidence that in both examples the semilattice order is a total order. However, we are
	not going into further detail in this place. An in depth investigation of the structure of midpoint cancellative convex
	semilattices is subject of our forthcoming work \cite{sokolova.woracek:cslmic-PRE}.
\end{remark}

\begin{definition}
\label{H30}
	Denote by $\bb W$ the class of all convex semilattices that are isomorphic to a subalgebra of a direct product of
	midpoint cancellative convex semilattices. 
\end{definition}

\noindent
In other words, $\bb W$ consists of those convex semilattices that can be embedded into a direct product with midpoint cancellative factors.
We note that $\bb W$ is closed under isomorphic images, subalgebras, and products.

Membership in $\bb W$ can be characterised also differently, namely via homomorphisms. That is actually a general fact, and uses the
following notion.
Let $X$ be a nonempty set, $\mc M$ a class of nonempty sets, and for each $M\in\mc M$ let $F_M\subseteq M^X$. 
We say that the class $\bigcup\{F_M\DS M\in\mc M\}$ is \emph{point separating} (or \emph{jointly injective}) on $X$, if 
\[
	\forall x,y\in X,x\neq y\DQ\exists M\in\mc M,f\in F_M\DP f(x)\neq f(y)
\]

\begin{lemma}
\label{H21}
	Let $X$ be a convex semilattice. Then $X$ can be embedded into a product with midpoint cancellative factors, if
	and only if the class
	\begin{equation}
	\label{H17}
		\bigcup\Big\{\Hom(X,B)\DSB B\text{ is a midpoint cancellative convex semilattice}\Big\}
	\end{equation}
	is point separating. 
\end{lemma}
\begin{proof}
	To see necessity, consider a product $Y\DE\prod_{i\in I}B_i$ of midpoint cancellative convex semilattices. 
	The canonical projections $\pi_i\DF Y\to B_i$ are homomorphisms, and the family of projections $\{\pi_i\}$ where 
	$i\in I$ is point separating on $Y$. It follows that \cref{H17} is point separating on $Y$, and thus on every 
	algebra isomorphic to a subalgebra of $Y$.

	For sufficiency assume that \cref{H17} is point separating. For each $x,y\in X$, $x\neq y$, choose a midpoint
	cancellative convex semilattice $B_{x,y}$ and $f_{x,y}\in\Hom(X,B_{x,y})$ with $f_{x,y}(x)\neq f_{x,y}(y)$. 
	Then the map 
	\[
		\left\{
		\begin{array}{rcl}
			\bb X & \to & \prod_{\substack{x,y\in X\\ x\neq y}}B_{x,y}
			\\[4mm]
			z & \mapsto & (f_{x,y}(z))_{\substack{x,y\in X\\ x\neq y}}
		\end{array}
		\right.
	\]
	is an injective homomorphism.
\end{proof}

\noindent
In the next lemma we exhibit a broad class of examples of convex semilattices belonging to $\bb W$. 
We point out that the lemma applies, in particular, to every free convex semilattice.
Recall here that a convex algebra $\langle X,+_p\rangle$ is called \emph{cancellative}, if 
\[
	\forall x,y,z\in X\DP\big(\exists p\in(0,1)\DP x+_pz=y+_pz\big)\Rightarrow x=y,
\]
and that $\langle X,+_p\rangle$ is cancellative if and only if it can be embedded homomorphically into some vector space over
the scalar field $\bb R$ (this is a classical fact going back to \cite{stone:1949,kneser:1952}).

\begin{lemma}
\label{H24}
	Let $\langle X,+_p\rangle$ be a cancellative convex algebra. Then $\Moewe X\in\bb W$.
\end{lemma}
\begin{proof}
	Choose a vector space $V$ over $\bb R$ and an injective convex algebra homomorphism $\iota\DF X\to V$.
	Then $\Moewe X$ is isomorphic to a subalgebra of $\Moewe V$. 
	By the Hahn-Banach separation theorem (e.g.\ \cite[Chp.IV,3.11\,Corollary]{conway:1990} applied with the weak topology on $V$
	induced by the algebraic dual $V^*$) we have 
	\[
		\forall A\in\Moewe V\DP 
		A=\bigcap_{f\in V^*}f^{-1}\Big(\big(-\infty,\max_{x\in A}f(x)\big]\Big)
		.
	\]
	The \emph{support functionals}
	\[
		\sigma_f\DF\left\{
		\begin{array}{rcl}
			\Moewe V & \to & \bb R
			\\
			A & \mapsto & \max_{x\in A}f(x)
		\end{array}
		\right.
	\]
	are convex semilattice homomorphisms, and by the above formula the family $\{\sigma_f\DS f\in V^*\}$ is point separating.
	Now refer to \Cref{H22}(ii) and \Cref{H21}.
\end{proof}

\subsection{The sufficiency result}

\begin{theorem}
\label{H4}
	Let $\langle X,+_p,\oplus\rangle$ be a convex semilattice in the class $\bb W$, let $I,J$ be nonempty sets, 
	and $\varphi\DF I\times J\to X$. If $\varphi$ satisfies \cref{H3} and \cref{H2}, then there exists a homomorphism 
	$\Phi\DF\mc CI\times\mc CJ\to X$ with $\Phi|_{I\times J}=\varphi$.
\end{theorem}

\noindent
The proof of the theorem relies on a trick, which shows how midpoint cancellativity enters.

\begin{lemma}
\label{H5}
	Let $B$ be a midpoint cancellative convex semilattice. 
	Let $I,J$ be nonempty sets, and $c_{ij}\in B$ for $i\in I,j\in J$.
	Then 
	\begin{equation}
	\label{H32}
		\forall i,i'\in I,j,j'\in J\DP
		c_{ij}+_{\frac 12}c_{i'j'}=c_{ij'}+_{\frac 12}c_{i'j}\wedge
		c_{ij}\oplus c_{i'j'}=c_{ij'}\oplus c_{i'j}
	\end{equation}
	if and only if
	\begin{equation}
	\label{H54}
		\Big(\forall i\in I\DQ\forall j,j'\in J\DP c_{ij}=c_{ij'}\Big)\vee
		\Big(\forall j\in J\DQ\forall i,i'\in I\DP c_{ij}=c_{i'j}\Big)
	\end{equation}
	The latter condition may be phrased as ``\/the matrix $(c_{ij})_{\substack{i\in I\\ j\in J}}$ 
	has constant rows or constant columns''.
\end{lemma}
\begin{proof}
	For the proof of the forward implication assume that \cref{H32} holds. 
	Assume the first bracket in \cref{H54} is false, so that there exist 
	$i_0\in I$ and $j_0,j_0'\in J$ such that $c_{i_0j_0}\neq c_{i_0j_0'}$. Let $j\in J$, then 
	\[
		c_{i_0j}\neq c_{i_0j_0}\vee c_{i_0j}\neq c_{i_0j_0'}.
	\]
	Consider the case that $c_{i_0j}\neq c_{i_0j_0}$. Let $i\in I$, then 
	\begin{equation}
	\label{H27}
		c_{i_0j_0}+_{\frac 12}c_{ij}=c_{i_0j}+_{\frac 12}c_{ij_0}\wedge
		c_{i_0j_0}\oplus c_{ij}=c_{i_0j}\oplus c_{ij_0}
		.
	\end{equation}
	Now midpoint cancellativity implies that 
	\[
		\big[c_{i_0j_0}=c_{i_0j}\wedge c_{ij}=c_{ij_0}\big]\vee
		\big[c_{i_0j_0}=c_{ij_0}\wedge c_{ij}=c_{i_0j}\big]
		.
	\]
	The first bracket is false, and we conclude that $c_{ij}=c_{i_0j}$. 

	If $c_{i_0j}\neq c_{i_0j_0'}$, we use 
	\[
		c_{i_0j_0'}+_{\frac 12}c_{ij}=c_{i_0j}+_{\frac 12}c_{ij_0'}\wedge
		c_{i_0j_0'}\oplus c_{ij}=c_{i_0j}\oplus c_{ij_0'}
	\]
	instead of \cref{H27}, and conclude again that $c_{ij}=c_{i_0j}$. Thus the second bracket in \cref{H54} holds true.

	The backwards implication is obvious.
\end{proof}

\begin{proof}[Proof of \Cref{H4}]
	Assume we are given nonempty sets $I,J$ and a map $\varphi$ subject to \cref{H3} and \cref{H2}. Recall that 
	$\Psi\DF\mc C(I\times J)\to X$ is the homomorphism with $\Psi(\{e_{ij}\})=\varphi(i,j)$, and 
	$\Lambda\DF\mc C(I\times J)\to\mc CI\times\mc CJ$ is the homomorphism with $\Lambda(\{e_{ij}\})=(\{e_i\},\{e_j\})$. 
	We are going to show that $\ker\Lambda\subseteq\ker\Psi$, which implies existence of $\Phi$.

	First consider the case that $X$ is midpoint cancellative. \Cref{H5} applied with the elements 
	$c_{ij}\DE\varphi(i,j)$ yields that the matrix $(\varphi(i,j))_{\substack{i\in I\\ j\in J}}$ 
	has constant rows or constant columns. Assume w.l.o.g.\ 
	that it has constant rows, i.e., $\varphi(i,j)$ is independent of $j$, and set 
	\[
		c_i\DE\varphi(i,j),\quad i\in I,j\in J
		.
	\]
	Let $\alpha\DF\mc CI\to X$ be the homomorphism with $\alpha(\{e_i\})=c_i$, and denote by 
	$\pi\DF\mc CI\times\mc CJ\to\mc CI$ the left projection. Then 
	\[
		\Psi\big(\{e_{ij}\}\big)=
		\varphi(i,j)=c_i=\alpha(\{e_i\})
		=(\alpha\circ\pi)\big((\{e_i\},\{e_j\})\big)
		=(\alpha\circ\pi\circ\Lambda)\big(\{e_{ij}\}\big)
		.
	\]
	Both, $\Psi$ and $\alpha\circ\pi\circ\Lambda$ are homomorphisms, and it follows that 
	$\Psi=\alpha\circ\pi\circ\Lambda$. In particular, therefore, $\ker\Lambda\subseteq\ker\Psi$.

	Now consider $X\in\bb W$. Let $B$ be a midpoint cancellative convex semilattice and $f\in\Hom(X,B)$.
	Since $f$ is a homomorphism, the map $f\circ\varphi$ satisfies \cref{H3} and \cref{H2}. 
	Its unique extension to $\Hom(\mc C(I\times J),B)$ is $f\circ\Psi$. By the already proved case, we have 
	\[
		\ker\Lambda\subseteq\ker(f\circ\Psi).
	\]
	Since the family \cref{H17} is point separating, it follows that $\ker\Lambda\subseteq\ker\Psi$. In fact, 
	let $(x,y)\notin\ker\Psi$, so that $\Psi(x)\neq\Psi(y)$. Choose $B$ and $f\in\Hom(X,B)$ such that 
	$f(\Psi(x))\neq f(\Psi(y))$. Then $(x,y)\notin\ker(f\circ\Psi)$, and hence $(x,y)\notin\ker\Lambda$.
\end{proof}

\section{Conclusions}

We investigate the property whether a map $\varphi$ from $I\times J$ to an algebra $X$ extends to a homomorphism from 
$\mc FI\times\mc FJ$ to $X$ (where $\mc FS$ denotes the free algebra with basis $S$). If the class of algebras contains commutative
binary operations, then a simple and explicit necessary condition for existence of such an extension follows. This condition happens to be
sufficient for semilattices and convex algebras, but not for convex semilattices. 
We show that for each two sets $I,J$ with at least to elements,
there exists a convex semilattice $X$ and a map $\varphi\DF I\times J\to X$ that satisfies the necessary condition but has no
homomorphic extension to the product. However, we also show that if only range spaces $X$ from a certain subclass of convex
semilattices are admitted, then the necessary condition is also sufficient.

%%%%%%%%%%%%%%%%%%%%%%%%%%%%%%%%%%%%%%%%%%%%%%%%%%%%%%%%%%%%%%%%%%%%%%

\nocite{*}
\bibliographystyle{fundam}
\bibliography{citations}

%%%%%%%%%%%%%%%%%%%%%%%%%%%%%%%%%%%%%%%%%%%%%%%%%%%%%%%%%%%%%%%%%%%%%%

\end{document}